\documentclass[letterpaper, 10 pt, conference]{ieeeconf}  

\IEEEoverridecommandlockouts                              

\usepackage{graphics} 
\usepackage{algorithm}
\usepackage{algpseudocode}
\usepackage{epsfig} 
\usepackage{mathptmx} 
\usepackage{amsmath} 
\usepackage{amssymb}  

\title{\LARGE \bf
AutoExp: A multidisciplinary, multi-sensor framework to evaluate human activities in self-driving cars}

\author{Carlos Crispim-Junior$^{1}$, Romain Guesdon$^{1}$, Christophe Jallais$^{2}$,
Florent Laroche$^{3}$,\\
Stephanie Souche-Le Corvec$^{3}$, Laure Tougne Rodet$^{1}$
\thanks{*This work was supported by French Region Auvergne-Rhône-Alpes}
\thanks{$^{1}$CFCJ, RG and LTR are with Univ Lyon, Lyon 2, LIRIS UMR 5205, F-69676, Lyon, France
        {\tt\small \{carlos.crispim-junior, romain.guesdon, laure.tougne\}@liris.cnrs.fr}}%
\thanks{$^{2}$CJ is with TS2-LESCOT, Univ Gustave Eiffel, IFSTTAR, Univ Lyon, F-69675 Lyon, France
        {\tt\small christophe.jallais@univ-eiffel.fr}}
\thanks{$^{3}$FL and SSL are with Univ Lyon, Université Lyon 2, LAET, F‐69007, LYON, France{\tt\small \{florent.laroche, stephanie.souche\}@cnrs.fr}}%
}

\begin{document}
\maketitle
\thispagestyle{empty}
\pagestyle{empty}

\begin{abstract}

The adoption of self-driving cars will certainly revolutionize our lives, even though they may take more time to become fully autonomous than initially predicted. The first vehicles are already present in certain cities of the world, as part of experimental robot-taxi services. However, most existing studies focus on the navigation part of such vehicles. We currently miss methods, datasets, and studies to assess the in-cabin human component of the adoption of such technology in real-world conditions. This paper proposes an experimental framework to study the activities of occupants of self-driving cars using a multidisciplinary approach (computer vision associated with human and social sciences), particularly non-driving related activities. The framework is composed of an experimentation scenario, and a data acquisition module. We seek firstly to capture real-world data about the usage of the vehicle in the nearest possible, real-world conditions, and secondly to create a dataset containing in-cabin human activities to foster the development and evaluation of computer vision algorithms. The acquisition module records multiple views of the front seats of the vehicle (Intel RGB-D and GoPro cameras); in addition to survey data about the internal states and attitudes of participants towards this type of vehicle before, during, and after the experimentation. We evaluated the proposed framework with the realization of real-world experimentation with 30 participants (1 hour each) to study the acceptance of SDCs of SAE level 4. 
\end{abstract}

\section{Introduction}

The adoption of self-driving cars will certainly revolutionize our lives. The first vehicles are already present in certain cities of the world, as part of continuous experimentations of robot taxis services. For instance, WAYMO proposes a robot taxi services in Phoenix, San Francisco and Los Angeles in USA. Other companies, such as Baidu, DeepRoute.AI, and Cruise also propose similar services.

Many studies have been devoted to advance the technological components linked to the navigation of the vehicle, from driving technology in different weather conditions, co-habitation with other road actors (pedestrians, bicycles, other vehicles)\cite{Parkin2022}, up to decision making in edge situations. For instance, what should the machine do in case of an avoidable car accident?

Even though the development and improvement of the navigation technology remains of great importance, the human (or vehicle's occupants) component of this equation remains lesser explored, particularly non-driving related activities.

We refer here to the activities we are likely to carry out when released from the task of driving (vehicles with SAE\footnote{Society of Automotive Engineers} level 4). These activities will likely have an impact on the way we spend time in the cockpit and consequently on the way the industry will design them in the future. 

Prior to recent advances in SDC development, most of existing work on human activity analysis in vehicle's cockpit was focused on driver monitoring for safety applications (detection of drowsiness, distraction from the road) \cite{DIAZCHITO201698}\cite{Weng2017}\cite{Roth2019}\cite{Ohn-Bar2013}\cite{Ahlstrom2013}. With the development of SDCs, researchers have shifted their focus to mechanisms for human-machine cooperative driving. For instance, in SDCs of SAE level 3, researchers have studied situations where the driver is asked to take back control of the vehicle, and seek to identify the activities that could compromise the change of responsability between the driver and the vehicle.

We current lack studies and the associated analytical tools that could help us analyze and anticipate how we will spend time in future SDCs of SAE level 4 (\textit{i.e.}, the non-driving related activities) in real-world conditions. The postures and activities that we will adopt and carry out during the usage of such types of SDCs are likely to affect our security and road safety in case of an accident, as well as how  we value (or invest) our time in the vehicle. 

We propose a multidisciplinary, multi-sensor framework to acquire data of human activities in a vehicle cockpit adapted to real-world conditions, named AutoExp. The framework acquires multidisciplinary data about the interior of the vehicle cockpit: from visual information with cameras up to subjective measurements about the occupants' internal states and their attitude towards this kind of vehicle.

\begin{figure*}[htb!]
\centering
\includegraphics[scale=0.75]{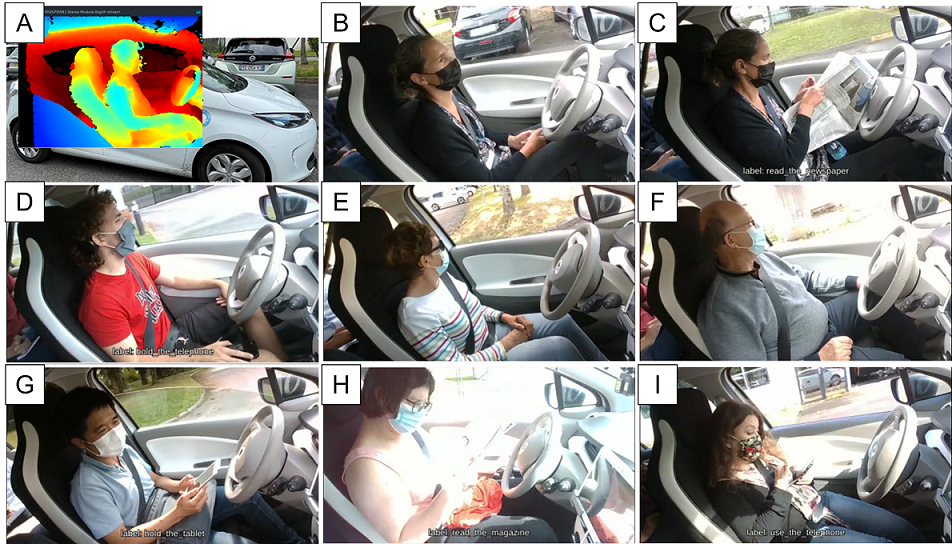}
\caption{Examples of recordings present in the AutoBehave dataset: A) Depth image of the cockpit environment of a Renault Zoe car programmed to behave like as a self-driving car with a SAE level of 4 (high automation). B-C) Images of a participant during the free and guided scenarios, respectively. D-I) Examples of different light conditions, variety of participants, as well as their usage of the time during the experiment with the SDC.}
\label{fig:dataset_images}
\end{figure*}

The proposed framework was used to carry out a real-world experimentation with 30 participants (1 hour each) testing a vehicle programmed to behave as an SDC of SAE level 4 (Fig. \ref{fig:dataset_images}). This experiment produced the AutoBehave dataset, a multi-sensor, multidisciplinary dataset for the study of non-driving related activities in SDCs, fully recorded in real-world conditions (in a vehicle and outside laboratory). The AutoBehave dataset depicts the activities of the occupants of an SDC, in as much naturistic driving conditions as possible, and by consequence, enable the development of tools to study the usage and the acceptance of such vehicles.

The remainder of this paper is organized as follows: Section \ref{secRW} presents and discusses the state of the art, Section \ref{secProp} introduces the proposed framework, Section \ref{secExp} details the experimentation we carried out, and Section \ref{secRD} presents preliminary results and discusses the lessons learned during the realization of the AutoBehave experimentation. Section \ref{secCFW} summarizes our contributions and shed light on future work.

\section{Related work}
\label{secRW}

Studies about consumer self-driving cars (SAE level 3 and 4) in human and social sciences, such as cognitive sciences and the economy of transport, are usually based on declarative surveys, that are realized online or in car cockpit simulators \cite{harb_what_2021}\cite{malokin_how_2019}. Even though studies about vehicles of SAE level 3 are beginning to appear in real world conditions, studies targeting consumer vehicles with the automation level 4, and that are carried out outside a laboratory, are much lesser developed.

Several tasks and datasets have been created to develop methods to monitor in-cabin human activity based in computer vision. Tasks range from the pose estimation of head, body and gaze \cite{Ahlstrom2013}\cite{DIAZCHITO201698}\cite{Guesdon_2021_ICCV}, up to the recognition of gestures \cite{Ohn-Bar2013}, driver states (\textit{e.g.}, drowsiness \cite{Weng2017}, distraction \cite{McDonald2020}), and actions
\cite{Pakdamanian2021}\cite{martin_driveact_2019}\cite{Ortega2020}. Datasets can be categorized based on their realism (synthetic versus real-world data) and on the target task (human head and body pose estimation \cite{Roth2019}\cite{Guesdon_2021_ICCV}\cite{guesdon_2023_VISAPP}, gesture and action understanding\cite{Ortega2020}\cite{da_cruz_sviro_2020}\cite{martin_driveact_2019}\cite{katrolia_ticam_2021}).

Synthetic data generation has become one of the driving forces on the development of activity monitoring systems using deep neural networks. Datasets are generally created using 3D scene models designed with tools, such as Blender\footnote{https://www.blender.org/}, Unity\footnote{https://unity.com/} and Unreal Engine\footnote{https://www.unrealengine.com}. For instance, SVIRO \cite{da_cruz_sviro_2020} is a dataset containing exclusively synthetic images of the interior of a car, more specifically the back seat. It offers several different sensor modalities, from RGB to infrared and scene depth images. It targets the evaluation of algorithms for object detection, human skeleton prediction and semantic segmentation. 

TiCAM dataset \cite{katrolia_ticam_2021} contains both synthetic and real multi-modal images (RGB, depth and infrared) of a car cockpit interior. But, real images are acquired in a car simulator. The dataset provides groundtruth to evaluate tasks such as 2D and 3D object detection, semantic segmentation, and gesture and action recognition. Action classes mostly consist in body movement (look left/right while driving, lean forward, turn left/right), however a few non-driving related actions are present, such as read paper ou book, use telephone, etc. Its main goal is to evaluate algorithms on the problem of domain adaptation: training a method on a source domain containing a large quantity of data (\textit{e.g.}, synthetic data), and later adapt the model to a target domain, where we generally possess a limited quantity of data (\textit{e.g.} real images). 

Guesdon \textit{et al.} \cite{guesdon_2023_VISAPP} proposed a large dataset (200~000 frames) containing synthetic images of 100 people in different poses inside of 7 different models of consumer vehicles. Unfortunately, this dataset focus exclusively on human pose related task. For instance, in human pose transfer, a method seeks to learn how to generate an image B from an image A of a person, for which the method only knows the image A, the estimated body keypoints of the person's body in A, and the target body keypoints of the human pose we expected to have in image B.

Very few datasets contain real images of a variety of people inside of a consumer vehicle in real world conditions. For instance, DriPE dataset \cite{Guesdon_2021_ICCV} contains 10~000 images of 19 people in different consumer vehicles (SAE level 0/1). It depicts several illuminations conditions from scene images records across day, night conditions included. However, it focuses exclusively on the tasks of pose estimation.

The DMD dataset (41 hours of RGB, depth and infrared videos recordings) \cite{Ortega2020}, and its subset dBehavior, focuses on the monitoring of driver's attention states and actions related to driving in vehicles of SAE levels 2 and 3. It contains recordings realized in a simulator and in a real vehicle. The dataset is composed of 37 participants (27 male, 10 female), where 10 worn glasses. Recordings capture 3 points of views (front face, hands, and a lateral view of the driver) using cameras from Intel RealSense D400 series. Examples of actions are: adjust radio, drive safe, hair make up, drinking, reach behind, talk left/right, talk to passenger, text left / right, reach side, hands free, hands free, switch gear. 

The Drive\&Act dataset \cite{martin_driveact_2019} is the only dataset containing real images of people carrying out non-driving related actions (simulation of a SAE level 4 vehicle). The recordings are made with RGB-D (Kinect for XBOX One video game console) and infrared cameras. The dataset contains 15 participants (11 male/4 female) with action classes hierarchically organized into categories, from atomic action units up to fine-grained activities. Atomic action term includes objects (\textit{e.g.}, bottle, laptop, \textit{etc.}), locations (\textit{e.g.}, driver door, right-back seat) and simple actions (\textit{e.t.}, reaching for, retracting, \textit{etc.}). Fine grained actions refer to more complex actions, such as drinking, eating, and opening the door. However, the entire dataset is recorded in an indoor driving simulator built with the cockpit of an Audi A3 car. 

The closest datasets to our proposal are DMD-dBehavior and DriveAndAct. DMD provides recordings of drivers in real-world conditions, but actions are mostly driving-related, since it focuses in driver assistance applications for vehicles of SAE levels 2 and 3. Drive\&Act dataset contains non-driving related actions, as it seeks to foster the evaluation of methods for activity analysis in future vehicles of SAE level 3 and 4. However, recordings are made in a cockpit simulator, limiting the realism of the acquired scene. 
Since both methods focus on evaluating methods for in-cabin monitoring, their experimental protocol is mostly described in terms of the sensor setup, and ignores other aspects we address here, such as the experimental scenarios we adopted to study the future usages of this kind of vehicle. 

In summary, the proposed framework novelties are the following: 1) an experimentation scenario to guide both the study of non-driving related activities of occupants of an SDC and the acquisition of varied examples of daily activities in a cockpit vehicle; 2) a sensor setup composed of a multiview array of cameras and surveys to monitor the activities in the cockpit; 3)
multi-disciplinary, multi-sensor dataset, namely AutoBehave, of comparable size to Drive\&Act and dBehavior/DMD datasets and Drive\&Act but fully recorded in an vehicle in outdoor environment. 

AutoBehave recordings are made in real-world conditions (a vehicle programmed to behave as a SDC SAE level 4, moving in the university parking and interacting with other road actors) and depict the activities self-engaged by people during their discovery of a vehicle programmed to behave as an SDC of SAE level 4.

\section{AutoExp framework}
\label{secProp}

\begin{figure}[h]
\centering
\includegraphics[scale=0.43]{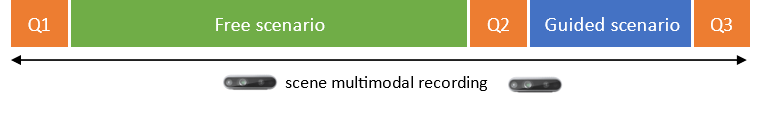}
\caption{Experimentation scenarios}
\label{fig:experimentation_scenario}
\end{figure}

\begin{figure}
\centering
\includegraphics[scale=0.25]{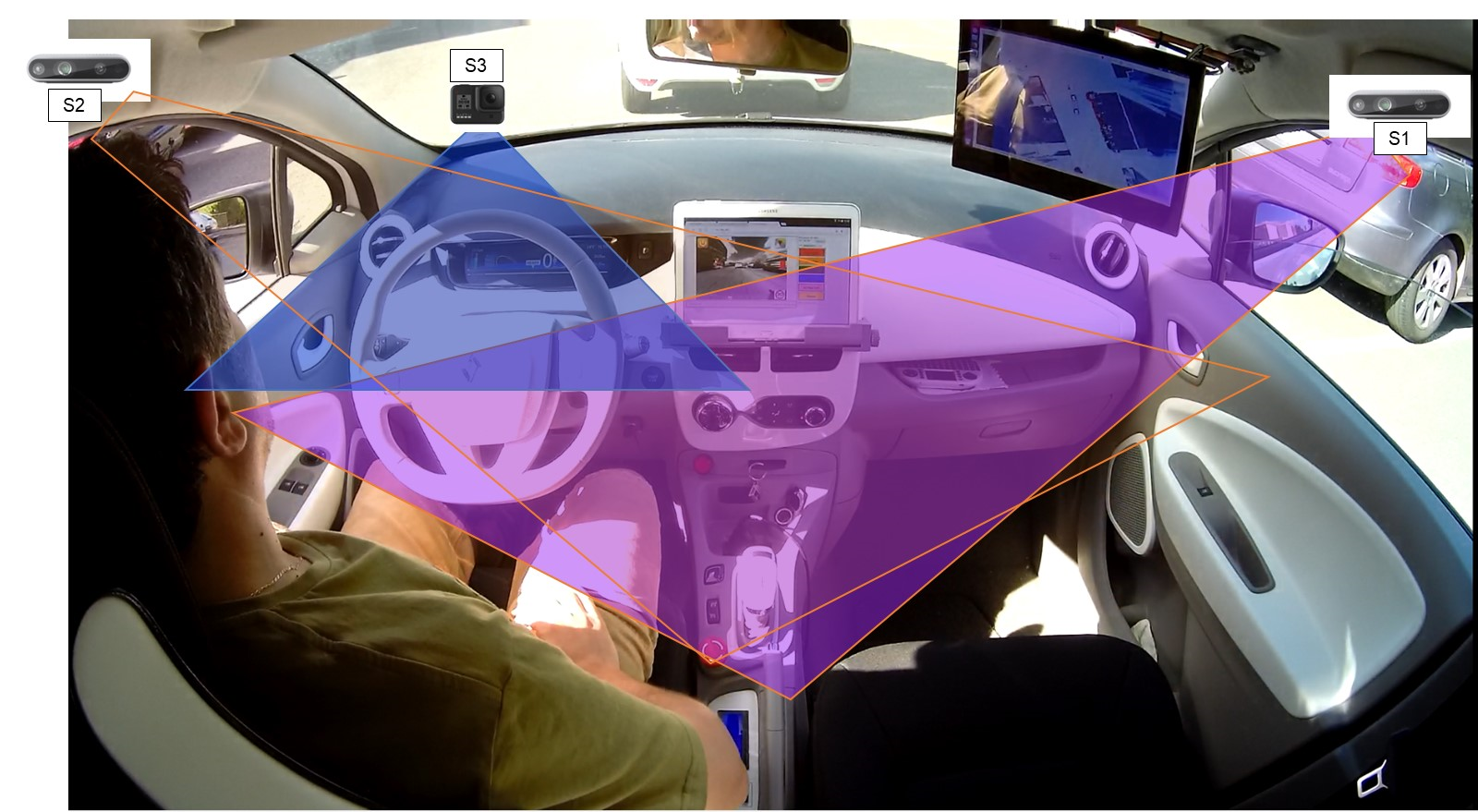}
\caption{Acquisition module setup: Sensors S1 and S2 capture a lateral view of the scene (Intel Realsense D435 cameras), and sensor S3 captures a front view of the upper body (GoPro model 8). Domain-specific data, such as internal states and attitudes towards the SDC, are acquired using surveys that are applied before, during and after vehicle usage.}
\label{fig:acquisition_setup}
\end{figure}

The AutoExp framework is composed of an experimentation scenario (Fig.\ref{fig:experimentation_scenario}), and an acquisition module (Fig. \ref{fig:acquisition_setup}). The experimentation scenario is composed of two parts: the free scenario and the guided (or data collection) scenario. Study participants are asked to fill a survey before and after each scenario.

The free scenario seeks to capture natural (non-guided) actions of the occupants of a vehicle during their discovry of the SDC. Participants are instructed to behave as they feel like in this scenario. They can carry out actions with objects they have brought with them, actions using the set of objects put at their disposal, or simply watch the experiment unfold idly.

The guided scenario takes place after the free scenario and it consists in the realization of a series of actions using the objects put at the  disposal of the participants. 
In this scenario, it is the experimenter that asks participants to begin or end the realization of an action, generally accomplished with the use of an object. The instructions are brief, and it is up to the participants to choose how to carry action. Examples of action instructions are: play with the tablet, read a magazine, or drink. The experimenter asks participants to change activities randomly. The objective of this scenario is to build a dataset of examples of human actions in a cockpit.

The experimental scenario is devised in an two-step fashion to be capable of both studying the natural behaviors of participants inside of a vehicle cockpit of SAE level 4, and also acquire a varied set of examples of people activities in the corresponding vehicle cockpit for the development of novel methods for pose and action understanding in SDCs.

The acquisition module is a multi-sensor framework to record multiple views of the front seats of the vehicle (Fig. \ref{fig:acquisition_setup}). It is composed of two RGB-D sensors (Intel RealSense D435) and a GoPro video camera. RGB-D sensors are attached to the left and right top corners of the windshield and capture a lateral view of the body of people at the rear seats. The GoPro camera records a close view of the upper body of the person in the driver seat. 

The multi-sensor recording data is complemented with data from two surveys. The first survey, named IT, collects data about the internal states (\textit{i.e.}, emotions) of the participants  before, between scenarios, and after the use of the vehicle. We adopted a commonly used scale in several studies to assess the emotion level felt by participants \cite{Scherer2005}. It is composed of 10 different emotions (\textit{e.g.}, anxiety, fear, serenity, alertness, anger) and participants had to respond using a 5-point Likert scale.

The second survey, named AT, assesses the attitude of participants towards this type of vehicle before and after the experimentation. The  questionnaries consisted in the confrontation of the attitude towards the SDC (trust, safety, productivity, etc.) and expected activities during driving (reading, sleeping, watching the road, etc.) before and after the experiment. Consequently, the questionnaries were similar to assess the change of attitudes and use of time before and after experiencing a driving session in a real SDC.

\section{AutoBehave experiment}
\label{secExp}

This section describes how we used the AutoExp framework to realize an experimentation and acquire a dataset depicting the test of an SDC of SAE level 4 (high driving automation). We carried out a four-day long experiment in July 2021 using a Renault Zoe car (electric supermini urban model) in the parking of the Ecole Centrale de Nantes (ECN) in France. The vehicle was robotized by the LS2N-ARMEN laboratory to behave as a SDC of SAE level 4. The developed framework enabled us to capture a multi-disciplinary, multi-sensor view of the actions realized by people during the test of the vehicle. 

Participants occupied the vehicle's driver seat and were instructed to not interfere with the vehicle navigation. A safety driver was sat in the passenger front seat and  used a dedicated interface to take back control of the car in case of need. An experimenter occupied the back seat and had the goal of starting/ending the experiment and of choosing the action sequence to be carried out during the guided scenario.
Participants were instructed to not engage in conversation with the experiment or the safety driver, unless they wanted to stop the experimentation. 

All participants signed a consent form allowing the project consortium to use and share their  data and video recordings for research purposes.

\subsection{Vehicle experimental path}

\begin{figure}
\centering
\includegraphics[scale=0.7]{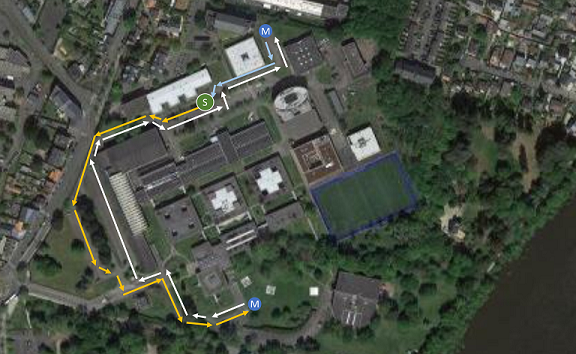}
\caption{Vehicle trajectory correspoding to one lap on the campus parking. S circle indicates starting/finishing point, M circles indicate turn around maneuvering points. The trajectory of the vehicle was dynamically adapted in response to the interaction with other road actors, from breaking/stopping up to overtaking parked vehicles. Source: Google Maps}
\label{fig:example_path}
\end{figure}

We defined a experimental path over the university parking (Fig.\ref{fig:example_path}). One lap in the defined path correspondeded to the vehicle traveling over the full path and coming back to the starting position (1.6 km distance in average). The lap path was recorded in advance and repeated to all participants. During the realization of the defined path, the vehicle adapted its trajectory using visual sensors (lidar, cameras) to overcome fixed obstacles (parked vehicles), and 
handle interactions with other road actors (pedestrians, bicycles and other vehicles). The road actors shared the parking streets with the SDC, unawared of the experiment taking place. During each lap, the vehicle carried out at least two turn-around maneuvers, since at the time of the experiment the parking was not connected in a circuit manner. 

Each participant experimentation took five laps. The first three laps were used for the free scenario (4.8 km in average). Participants were asked to behave as if they were in a daily travel from domicile to work.  This scenario aimed to collect data to evaluate their usage of the time in the cockpit and enable the study of their  acceptance of an SDC. The last two laps corresponded to the guided scenario (3.2 km in average). The experimenter instructed participants to carry out a series of actions using the objects put at their disposal. The order of guided activities, as well as their duration was chosen randomly.

\subsection{Scenarios' sequence and metaparameters}

The algorithm \ref{alg:exp_seq} summarizes the sequence of scenarios used for the AutoBehave experimetation with the chosen metaparameters. The free scenario lasted for 26 minutes in average, while the guided scenario lasted for 18 minutes in average. The total time per participant was close to 1 hour. The duration of scenarios varied due to interaction with other road actors, that sometimes required that the safety driver take back control of the vehicle. We applied the IT and AT survey questionnaires before starting the vehicle, and just after its use. The IT survey was also applied in between scenarios.

\begin{algorithm}
\caption{Experimentation algorithm}
\label{alg:exp_seq}
$ApplySurvey(IT, AT)$\\
$FreeScenario(3\ laps)$\\
$ApplySurvey(IT)$\\
$GuidedScenario (2\ laps)$\\
$ApplySurvey(IT,AT)$\\
\end{algorithm}

\section{Results and discussion}
\label{secRD}

This section presents the dataset we acquired with the AutoExp framework, the selected set of actions chosen for the guided scenario, and the preliminary analysis of the experiment.

\subsection{AutoBehave dataset}

We acquired a multidisciplinary, multi-sensor dataset composed of 29 people (18 men/11 women) carrying out actions in the context of a daily domicile-work travel during the test of an SDC of automation level 4 (Fig. \ref{fig:dataset_images}). The dataset is composed only of the participants for which we have valid video recordings and domain specific surveys.

\begin{figure}
\centering
\includegraphics[scale=0.4]{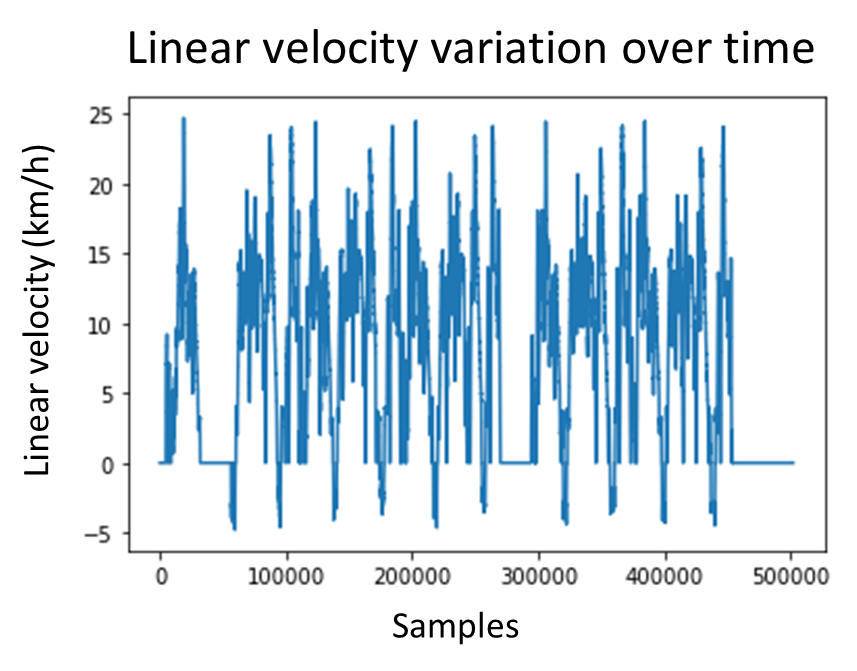}
\caption{Example of the variation of linear velocity (km/h) of the vehicle across the experiment for one participant. Negative values correspond to the vehicle running backward}
\label{fig:motion_graph}
\end{figure}

Participants were recruited based on their declared interest to test a novel technology during their participation in prior, online survey on people attitude towards SDCs, or in their response to recruitment announces made to the employees and students of the university. They present a variety of body sizes, human traits, ages, and education levels. Multi-sensor video recordings are 1-hour long and are accompanied by the domain-specific questionaries (IT, AT). The vehicle traveled for nearly eight kilometers per participant, with an maximum speed of 25 km/h. Total distance slightly varied across participants due to interactions with other road actors that may have required unexpected maneurvering events (e.g., to overcome improper parking). Fig. \ref{fig:motion_graph} illustrates the variations of linear velocity across the experiment for a participant. Experiment duration also varied  due to the time taken by certain participants to respond the questionnaires.

The acquired dataset succesfully depicts real-life situations from calm journeys up to abrupt interactions with other road actors (bycicles crossing path with the vehicle without clear warning, delivery trucks badly parked, \textit{etc}). The safety driver has taken back control of the vehicle in a few situations as a measure of safety, when other vehicles did not respect the distance between vehicles or overtake the SDC in an agressive manner.

The decomposition of the experiment into two scenarios worked well and enabled us to acquired a large volume of video data without interfering with the participants "simulation" of a daily home to work journey. We obtained nearly 30 hours of recordings, where at least 500 minutes consist in examples of actions people may realize in the cockpit of an SDC level 4. The recordings are also very rich in terms of real-life sources of noise during cockpit monitoring. For instance, we observed varied illumination changes and different types of actions interruptions, varing from those thriggered by natural vibrations of the vehicle's displacement until those related to abrupt stops caused by the road conditions.

\subsection{Action catalog}
\label{sec:action_catalog}

One of the objectives of AutoBehave experiment was to investigate the activities that people will freely engage during their free time in the cockpit of an SDC, when released from the task of driving. We equiped the vehicle with a experiment bag containing several objects involved in daily actions that people could do in an SDC. The action vocabulary adopted (Table \ref{tab:activity_classes}) was constructed based on the study of existing computer vision datasets (like Drive\&Act dataset) and activities envisaged by people when questioned by online surveys about future usages of SDC's cockpit time \cite{SoucheSSRN2022}.

\begin{table}[h]
\caption{Proposed activities}
\centering
\begin{tabular}{|l|l|}
\hline
\textbf{Object} 	&	 \textbf{Action} 	\\
\hline
water bottle 	&	 drink 	\\
telephone 	&	take a telephone call  	\\
telephone 	&	 make a telephone call  	\\
tablet 	&	 play a game in the tablet  	\\
magazine 	&	 read the magazine  	\\
newspaper 	&	 read the newspaper  	\\
work bag 	&	 search in the bag  	\\
telephone 	&	 listen to music  	\\
\hline
\end{tabular}			
\label{tab:activity_classes}
\end{table}

The experiment bag was available to participants during all scenarios. However, we observed that most participants only used it in the second scenario, when they were explicitly guided by the experimenter to carry out actions with the bag's objects. All participants realized all actions in the guided scenario.

\subsection{Preliminary behavioral analysis}

This section describes the preliminary analysis we carried out using the video recording of the AutoBehave dataset. In the free scenario, we observed that 9 of the 29 participants watched the experimentation idly. Similarly, 12 of the 29 participants interleaved the observation of the road with the action of "using of the telephone".
In a minor frequency, 3 participants spent their time reading a magazine or a book, and 2 recorded the road, sometimes resting. In the guided scenario, participants have carried out all actions asked without problem, enabling us to build a collection of video recordings about human actions in the cockpit of a SDC of level 4. We observed a few actions in the free scenario that are common in real life, but were not covered by the initially defined vocabulary. They were video recording the scene with a smartphone and making a video call. These actions were not frequent, but could be explained by the fact participants were discovering a new technology.

The comparison of action classes realized in the free and guided scenarios suggest participants could carry out the proposed actions in the vehicle cockpit without a problem. However, the majority of them has chosen not to do it in the free scenario. This absence of activities in the first scenario could be motivated by several factors, such as a need to be attentive to the vehicle navigation, the fact participants are focused on the discovery of this new type of technology, among other factors. Further analysis of the dataset with the acquired multidisciplinary data is necessary to understand the origin of the absence of activities in the free scenario. We also observed frequent human body activity in the free scenario that was related to body pose changes, hand gestures, or brief, simple actions. These observations suggest that methods for human body keypoint estimation and gesture recognition will also have a role to play in human activity analysis for the study of people's usage of time in future SDC cockpits.

The experiment decomposition into two scenarios has proved to be a successful approach. It enabled us to observe what actions people will naturally and freely engage in when using an SDC, while reserving a slot of time, the guided scenario, as a safeguard measure to acquire valuable data for further studies on methods for the automatic analysis of cockpit activity. In terms of the evaluation of the usage of time in the vehicle cockpit, the described results are promising, but remain an exploratory research in the topic. A larger cohort of participants is necessary to validate the  described observations, given the small sample-size of our dataset (29 particpants), with respect to other disciplines' requirements. Further development is also necessary to consider other factors, like the habituation of participants to the technology and the effect it may have on people's internal states and their will to engage in activities.

\section{Conclusion and Future work}
\label{secCFW}

We have proposed a multidisciplinary, multi-sensor framework to study the activities of the occupants of SDCs of SAE level 4. We demonstrated its usage with the acquisition of a multidisciplinary dataset depicting the non-driving related activities of people during the test of a vehicle programmed to behave like a SDC of SAE level 4 and traveling in an outdoor environment. The acquired dataset successfully depicts a large varied of real-life sources of noise both on sensor data and in action realization. The described results have show the interest of the proposed framework for the analysis of human activities in SDCs. Further experimentation is necessary to validate the experimental observations about the usage of vehicle time with a large cohort of participants. Future work will focus on the development of methods to analyze the acquired data automatically, in addition to the study of the acquired dataset with multidisciplinary tools.

We hope that the developed framework will serve as the basis for other projects' experiments in the field of autonomous vehicles and that it can support regulatory and societal studies about future usages of Intelligent Transport Systems, such as SDCs. The AutoBehave dataset will be released progressively under the form of computer vision tasks and multidisciplinary studies to foster the development of methods to monitor the cockpits of SDCs.

\addtolength{\textheight}{-12cm}   


\section*{Acknowledgment}

This work has been realized in the context of the AURA AutoBehave project 2019, a French project devoted to developing novel methods to evaluate the acceptability of autonomous vehicles using a multidisciplinary approach. We thank the Region Auvergne-Rhône-Alpes for funding this research.


\bibliography{itscrefs}
\bibliographystyle{plain}

\end{document}